\documentclass[10pt,conference]{IEEEtran}

\usepackage{amsmath}
\usepackage{amssymb}
\usepackage{graphicx}
\usepackage{float}
\usepackage{wrapfig}
\usepackage{epsfig}
\usepackage{psfrag}

\begin{document}

\title{On optimizing low SNR wireless networks using network coding}

\author{
\authorblockN{Mohit Thakur}
\authorblockA{Institute for communications engineering,\\
Technische Universit\"{a}t M\"{u}nchen,\\
80290, M\"{u}nchen, Germany.\\
Email: mohit.thakur@tum.de}
\and
\authorblockN{Muriel M\'{e}dard}
\authorblockA{Research Laboratory for Electronics,\\
MIT, Cambridge, MA, USA.\\
Email: medard@mit.edu}
}

\vspace{-6mm}
\maketitle
\vspace{-2mm}
\begin{abstract}
The rate optimization for wireless networks with low SNR is
investigated. While the capacity in the limit of disappearing SNR is
known to be linear for fading and non-fading channels, we study the
problem of operating in low SNR wireless network with given node
locations that use network coding over flows. The model we develop
for low SNR Gaussian broadcast channel and multiple access channel
respectively operates in a non-trivial feasible rate region. We show
that the problem reduces to the optimization of total network power
which can be casted as standard linear multi-commodity min-cost flow
program with no inherent combinatorially difficult structure when
network coding is used with non integer constraints (which is a
reasonable assumption). This is essentially due to the linearity of
the capacity with respect to vanishing SNR which helps avoid the
effect of interference for the degraded broadcast channel and
multiple access environment in consideration, respectively. We
propose a fully decentralized Primal-Dual Subgradient Algorithm for
achieving optimal rates on each subgraph (i.e. hyperarcs) of the
network to support the set of traffic demands (multicast/unicast
connections).

\emph{Index Terms -} Low SNR Gaussian broadcast channel, network
coding, rate optimization, Primal-Dual Subgradient Method.
\end{abstract}

\vspace{-2mm}
\section{INTRODUCTION}
\vspace{-2mm} Wideband fading channels have been studied since the
early 1960's. Kennedy showed that for the Rayleigh fading channel at
the infinite bandwidth limit, the capacity is similar to the
capacity of the infinite bandwidth AWGN channel with the same
average received power [1, 2]. The robustness of this result in the
case of with or without channel state information helps us model the
low SNR wideband wireless networks in a general manner. It should be
noted that when the band grows large, the transmitting power is
shared among large degrees of freedom. This results in smaller SNR
per degree of freedom. Using this as our underlying
information-theoretic assumption to approximate the capacity over a
link, we model the general traffic for this network and show that
the linearity of capacity for disappearing SNR makes for the
fundamental reason for simplicity in our model. Hence, we claim, it
is possible to do networking over such a model with simplistic and
essentially linear approach.

In the context of wideband multipath fading relay channel, it was
shown in [3] that in the non-coherent multipath fading relay
channel, the same lower bound on the rate can be achieved as in the
frequency division AWGN relay channel with the same received SNR, by
using a peaky binning scheme. In this paper, we use a relaying
scheme based on superposition coding. The rates achieved by peaky
binning [3] are higher than the rates achieved by the relaying
scheme that we propose. However, our relaying scheme has the
advantage to extend easily to large networks due to hyperarc
decomposability. We would like to mention here that number of
hyperarcs for Gaussian broadcast channel using superposition coding
is equal to $n$ for $n$ receivers, instead of $2^{n}$.


The traffic model we use is quite general. It is divided into two
classes: unicast and multicast (broadcast is considered as a special
case of multicast), where each pair of source and receiver group in
the network form a session for a particular class of traffic. The
problem of successfully establishing multicast connections in
wireline or wireless networks has been long thought to be
NP-Complete using arbitrary directed and undirected network models.
With the advent of network coding (ref., [4], [5], [6]), and in turn
breaking of the fluid model for data networks i.e. by performing
coding over incoming packets, this approach has been able to
intrinsically circumvent the combinatorial hardness of the multicast
flow problem. It was also shown that establishing minimum cost
multicast connections boils down to optimizing subgraph over coded
packet networks [7].

In this paper, we consider a low SNR wireless network with Gaussian
broadcast channel and MAC. The problem we consider is to optimize
the rates for a given set of demands that needs to be met by this
network. We show that this problem can be casted as a minimum cost
multicommodity flow problem with intra-session network coding.

This paper is organized as follows. Section II is composed of
general problem formulation. In section III we propose a
decentralized solution. We present our results in section IV and
finally, we mention concluding remarks in section V.

\section{SET-UP AND PROBLEM FORMULATION}
\vspace{-2mm} In this section we introduce the hypergraph models for
the low SNR Gaussian broadcast and multiple access channel.

\begin{figure}
   \begin{center}
    \includegraphics[width=0.47\textwidth]{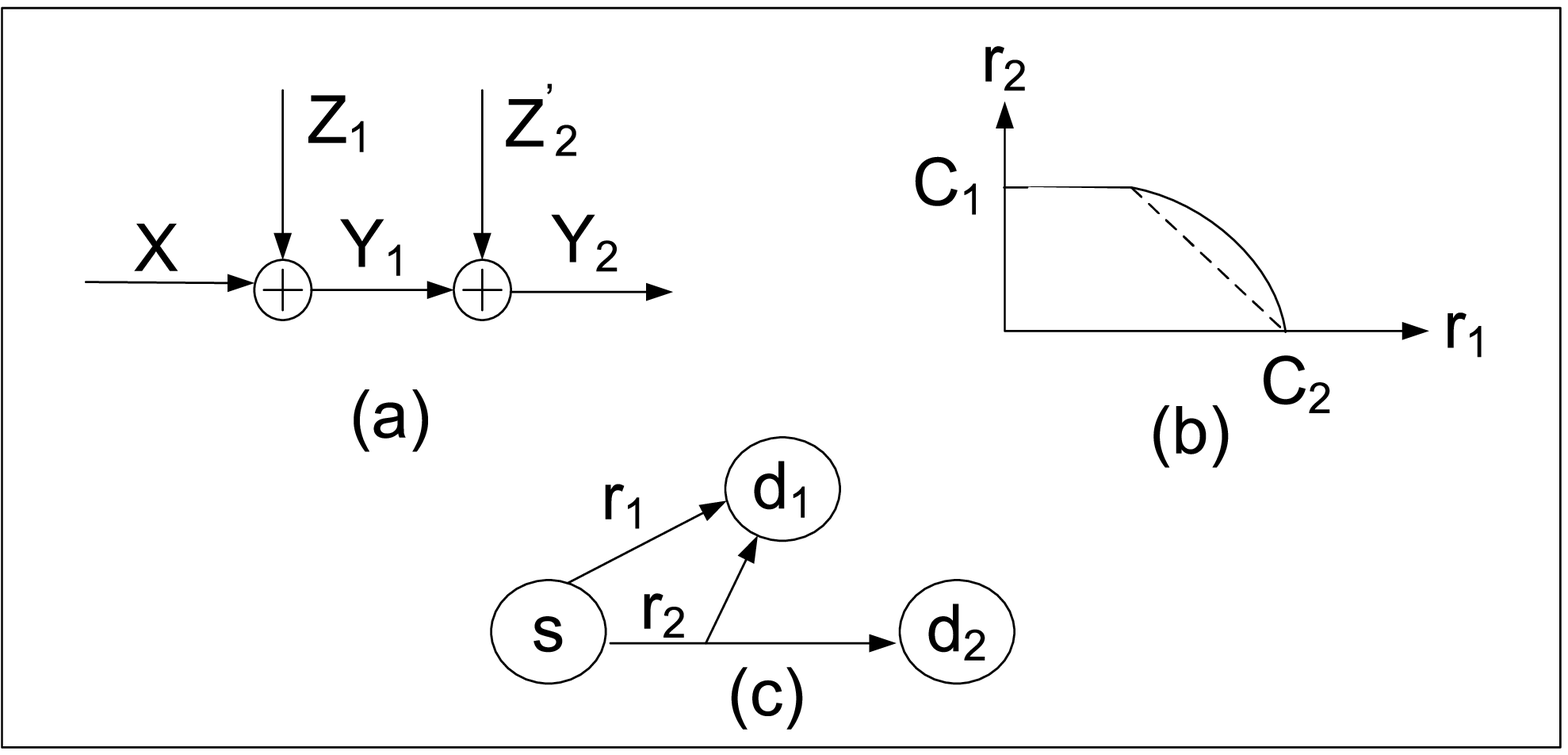}
\vspace{-4mm}
\end{center}
  \caption{(a): Two receiver physically degraded Gaussian broadcast channel with $Z_{1} \sim \mathcal{N}(0, N_{1})$ and $Z_{2}^{'} \sim \mathcal{N}(0, N_{2}-N_{1})$. (b): Rate region for the channel in (a), dotted line denotes the flatness of the rate region curve in the limit of vanishing SNR with $C_{1}$ and $C_{2}$ as max rates for each receiver respectively. (c): Decomposition into hyperarcs $\{(s,(d_{1})),(s,(d_{1}, d_{2}))\}$ with their common rates for the case in (a) with receivers $d_{1}$ and $d_{2}$ (corresponding to better and worse respectively).} \vspace{-6mm}
\label{Broadcast channel and hypergraph decomposition}
\vspace{-1mm}
\end{figure}

\textit{A. Low SNR physically degraded Gaussian broadcast channel [14].} \\
Consider a general wideband fading channel where the input waveform
is $\textbf{x}$ and the output waveform is $\textbf{y}$, the fading coefficient matrix is given by $\textbf{h}$ and
$\textbf{n}$ is the additive white noise. The channel is given by:
\begin{equation}
\textbf{y}= h \textbf{x} + \textbf{n}. \vspace{-1mm}
\end{equation}
The capacity of the channel, for both Gaussian
channels and fading channels increase sublinearly with the increase
in signal to noise ratio $(SNR)$ but in the low SNR regime the
capacity in the limit is linear in SNR for fading and non-fading
channels:
\vspace{-2mm}
\begin{equation}
\vspace{-2mm}
C(SNR) = SNR + o(SNR) (nats/s/Hz).
\end{equation}
Clearly at low SNR, the signal-to-noise ratio per degree of freedom
(SNR) approaches zero in the limit [2, 8, 9]. For such a case the
point to point capacity boils down to:
\begin{equation}
C_{AWGN}=\frac{Ph^{2}}{N_{0}}=lim_{W \rightarrow \infty} W log
(1+\frac{Ph^{2}}{WN_{0}}),
\end{equation}
where $h^{2}=\frac{1}{D^{\alpha}}$ and $D$ is the distance between
the transmitter and receiver. Let us now look at the standard model
of a single sender and two receivers with noise variances $N_{1}$
and $N_{2}$ respectively (ref. Fig 1(a)). The capacity region is
given by:
\begin{align}
r_{1} < C(\frac{\lambda_{1} P}{N_{1}}),  r_{2} < C(\frac{(\lambda_{2})
P}{\lambda_{1} P+N_{2}}).
\end{align}
where $C(x)=W(ln(1+x))$, $\lambda_{1}+\lambda_{2} =1$, $\lambda_{i}
\geq 0$ and $P$ is the total power (ref. Fig. 1(b)), the
transmission scheme is superposition coding [15].


The rate region defined in (4), when looked under the low SNR lens
comes across as a rather simpler picture.  For the power limited low
SNR regime, the effect of the power allocated for the better
receiver, as the contribution to the total noise experienced by the
worse receiver is negligible (ref. Fig. 1(b), for the rate region
for low SNR in the limit). So, for the low SNR physically degraded
Gaussian broadcast channel, the rate for the worst receiver can be
approximated as \vspace{-1mm}
\begin{equation}
r_{2} \lessapprox C(\frac{\lambda_{2} P}{N_{2}}).
\vspace{-1mm}
\end{equation}
Generalizing the same idea for the case of a given source $i$ with
power $P_{i}$ and $n$ receiver nodes, where the receiver set
$J=(1,...,n)$ can be broken into $n$ subsets as $J^{k}=(1, 2,
...,k)$ for $k \in [1, n]$ ordered in decreasing order of
reliability. The rate region defined for each hyperarc $(i, J^{k})$
in the low SNR limit is given as \vspace{-1mm}
\begin{align}
r_{iJ^{k}} &\lessapprox C(\frac{(\lambda_{k})P_{i}}{N_{k}}),  \forall k \in [1,n] \\
& \approx \frac{(\lambda_{k})P_{i}}{\parallel L_{i} - L_{k}
\parallel^{\alpha} N_{2}}, \forall k \in [1,n]. \vspace{-6mm}
\end{align}
where, $\displaystyle\sum_{k=1}^{n} \lambda_{k} \leq 1$, which when
combined appropriately gives the rate region of the set $J^{k}$. The
equation (7) comes from the fact that capacity is linear in the
limit of disappearing SNR, where $L_{i}$ for all $i \in [1,n]$ is
the location of the node and $\alpha$ is the loss exponent. We
formalize the above mentioned concepts and motivate our next
definition. Let $\lambda_{k} P_{i} = P_{iJ^{k}}$, $\forall k$.

\textit{Definition. 1: For a given sender $i$ with total power
$P_{i}$ and a receiver set $J=[1,K]$ in low SNR physically degraded
Gaussian broadcast channel, the set $J$ can be decomposed into $K$
hyperarcs where each hyperarc is defined as the connection from the
sender $i$ to the receiver set $J^{k}=[1,k]$, where $k \subseteq
[1,K]$.  The rate over each hyperarc is defined as $r_{iJ^{k}} =
r_{iJ^{k}}= \frac{P_{iJ^{k}}}{\parallel L_{i} - L_{k}
\parallel_{2}^{\alpha} N_{2}},$ where, $\displaystyle\sum_{k}
P_{iJ^{k}} \leq P_{i}$, $\forall k \in J$ and the set $J^{k}$ ranges
from best to worst receiver (ref. Fig 1 (c)).}

\vspace{2mm}
\textit{B. Interference issues in multiple access at low SNR.} \\
Now, let's consider the case of multiple access where more
than one node tries to access the channel at the given instance. Let
there be $U$ nodes in the system at an instance, and $u \subset U$
of them are trying to access the channel at this instance, if node
$i \in u$ intends to communicate with node $j \in U$  among others in $u$, the signal to interference and noise ratio
(SINR, denoted as $\mu_{ij}$) experienced at node $j$ is given
by: \vspace{-2mm}
\begin{equation}
 \mu_{ij} = \frac{\frac{P_{i}}{\|L_{i} - L_{j}\|^{\alpha}}}{W(N_{0} + \displaystyle\sum_{v \in u, v \neq i} \frac{P_{v}}{\|L_{v} - L_{j}\|^{\alpha} N_{0}})}.
\vspace{-2mm}
\end{equation}
Note that, since every node in $u$ is interested only in a common receiver, we allocate the whole power of the node over this single hyperarc, so $k=1$ and $P_{iJ^{1}}=P_{i}$ for every transmitter. But as we are operating in the low SNR regime, the intuition suggests that
the effect of the interference should be negligible. We straightforwardly include it in
our assumption, thus we define the rate (denoted with $R$) experienced at the receiver
$j$ as:
\vspace{-4mm}
\begin{align}
R_{ij} &= W ln \bigg(1 + \frac{\frac{P_{i}}{\|L_{i} - L_{j}\|^{\alpha}}}{W(N_{0} + \displaystyle\sum_{v \in u, v \neq i} \frac{P_{v}}{\|L_{v} - L_{j}\|^{\alpha}N_{0}})} \bigg) \\
& \approx W ln \big(1 + \frac{P_{i}}{W(\|L_{i} -
L_{j}\|^{\alpha} N_{0})} \big) \\
& \approx W\big(\frac{P_{i}}{W(\|L_{i} - L_{j}\|^{\alpha} N_{0})}
\big)=\big(\frac{P_{i}}{\|L_{i} - L_{j}\|^{\alpha} N_{0}} \big).
\end{align}
The approximation (10) comes from the fact that the contribution of
other signals being transmitted from other sources in the system
with low SNR channel to the interference is negligible and the
approximation (11) comes from the linearity of capacity in the limit
of disappearing SNR (ref. Fig 2(a) and 2(c)). In Fig. 2(b), we can
see that the SNR curve approaches the capacity curve in the limit,
corroborating our assumption that the SNR equals capacity in the
limit of disappearing SNR per degree of freedom.

\begin{figure}
   \begin{center}
    \includegraphics[width=0.47\textwidth]{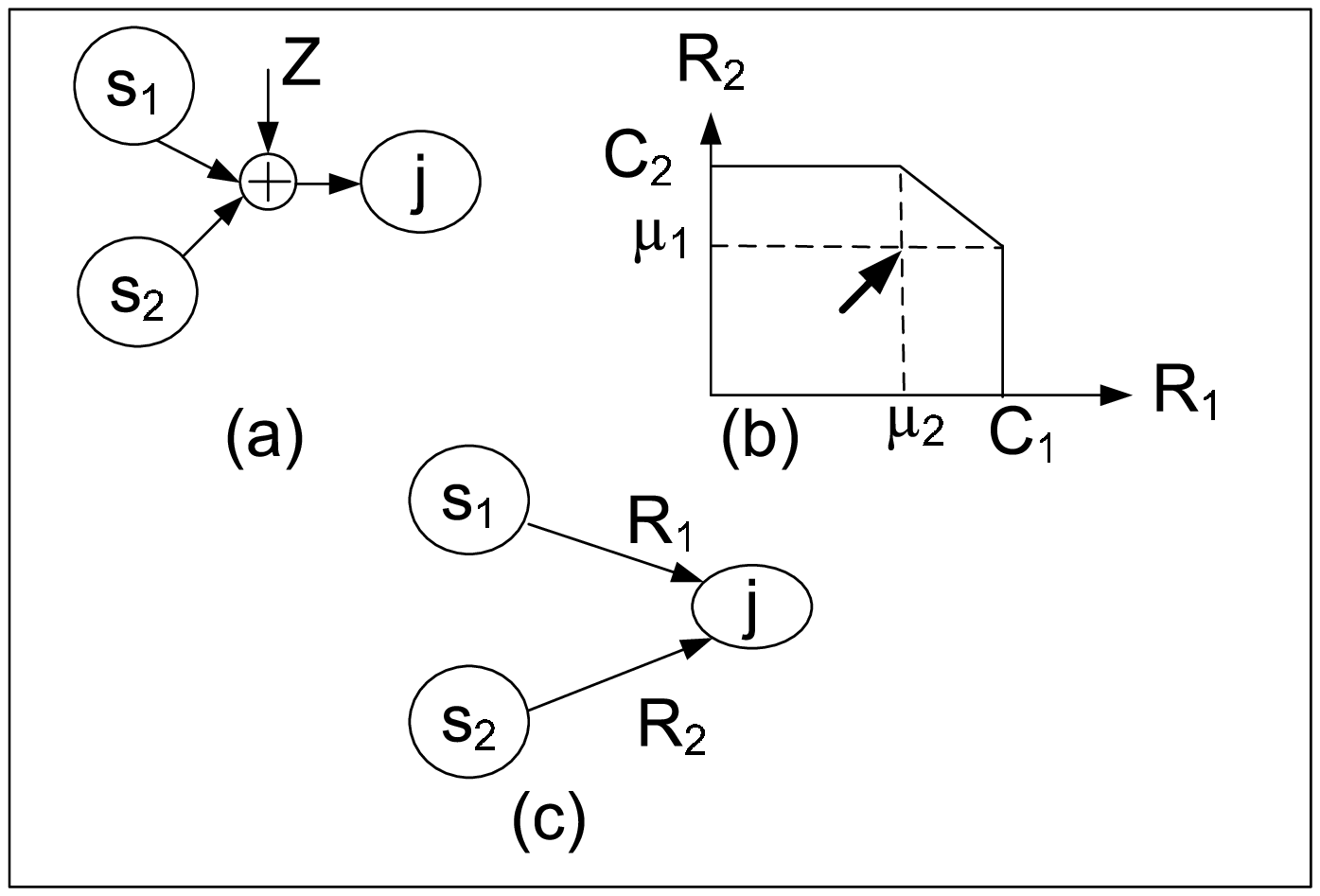}
\vspace{-3mm}
  \end{center}
  \vspace{-2mm}
  \caption{(a): Two sender case for the low SNR multiple access channel, where $Z \sim \mathcal{N}(0, N)$. (b): Rate region for case in (a), the dotted line denotes the respective SNR's $\mu_{1}$ and $\mu_{2}$ for two senders and the arrow shows that in limit of disappearing SNR, the SNR curve touches the capacity curve. (c): As the effect of interference is negligible, the case is (a) can be approximated as individual hyperarcs.}
\label{Multiple Access}
\vspace{-6mm}
\end{figure}

\vspace{2mm}
\textit{C. Low SNR network rate optimization.} \\
Let us represent the wireless network as a directed hypergraph
$\mathcal{G = (N,A)}$, where $\mathcal{N}$ is the set of nodes and
$\mathcal{A}$ is the set of hyperarcs, where each hyperarc emanates
from a node and a terminates at a group of nodes, which we also
refer to as the broadcast group of the hyperarc. Note that we
consider multicast in our multicommodity flow optimization model (as
opposed to only unicast), thanks to network coding.

It's important to note that the common rate associated with each
hyperarc  $\mathit{r_{iJ^{k}}}$, is the capacity of the hyperarc,
because this is the rate that can be guarantied to all the receivers
in this hyperarc. Also, $\mathit{r_{iJ^{k}}}$, is a nonnegative
function of the transmit power $\mathit{P_{iJ^{k}}}$ of the hyperarc
${(i,J^{k}})$. Now that we consider a network with more than one
sender, update of notations is required. For a sender $i \in
\mathcal{N}$, that is capable of reaching $k_{i} \in
(1_{i},..,K_{i})$ nodes, where each $k_{i} \in \mathcal{N}
\backslash i$, the $K$ hyperarcs are denoted by $(i,J^{k_{i}})$,
$\forall k_{i} \in (1_{i},..,K_{i})$.

Imagine a set of traffic demands where $m=1,...,M$ sessions need to
be established, each with $t_{m}=1,....,T_{m}$ set of receivers, in
a given wireless network that experiences low SNR and that is
represented by the hypergraph $\mathcal{G = (N,A)}$. We know from
the definition of hyperarc that a single node can lie on multiple
hyperarcs, therefore, we need a way to carefully count the incoming
information and outgoing information to apply the law of flow
conservation to the hypergraph and finally be able to cast the
problem as a flow optimization problem. For that, we define another
graph $\mathcal{G' = (N,A')}$, which is simply the equivalent
directed graph of $\mathcal{G = (N,A)}$ with arcs instead of
hyperarcs. This graph can be easily obtained by decomposing the
hypergraph appropriately. Let's define the term (ref. [7] for
detailed notation explanation): \vspace{-2mm}
\begin{equation}
x_{iJ_{l}^{k_{i}}} = \displaystyle\sum_{((i,J^{k_{i}}) \in \mathcal{A} | J^{k_{i}} \ni l)} x_{iJ^{k_{i}}}.
\vspace{-2mm}
\end{equation}
which simply describes the way to add all the flow entering a node
on all incoming hyperarcs, corresponding to the graph $\mathcal{G' =
(N,A')}$. Notice that $x_{iJ^{k_{i}}_{l}}$ is not the same as
$r_{iJ^{k_{i}}_{l}}$ defined in the previous section,
$x_{iJ_{l}^{k_{i}}}$ can be interpreted as the flow between $i$ and
receiver $l$ of the hyperarc $J_{k_{i}}$, and it cannot exceed the
common rate ($r_{iJ^{k_{i}}}$) associated to the hyperarc which is
also the hyperarc capacity, for each $l\in k_{i}$.

Let, $r_{iJ^{k_{i}}} = \frac{P_{iJ^{k_{i}}}^{k_{i}}}{\|L_{i} -
L_{k_{i}}\|^{\alpha}N_{0}}=\gamma_{iJ^{k_{i}}}P_{iJ^{k_{i}}}^{k_{i}}$.
Then, the minimum cost optimization problem for the low SNR network
can be formulated as:
\begin{align*}
minimize \displaystyle\sum_{(i,J^{k_{i}}) \in \mathcal{A}} P_{iJ^{k_{i}}}^{k_{i}}  &&\text{(A)}
\vspace{-4mm}
\end{align*}
subject to:
\vspace{-3mm}
\begin{gather}
\mathit{y_{iJ^{k_{i}}}(m)} \geq \displaystyle\max_{t_{m}}(x_{iJ^{k_{i}}}^{t_{m}}(m)), \forall (i,J^{k_{i}})\in \mathcal{A}, \forall m \\
z_{iJ^{k_{i}}}=\displaystyle\sum_{m=1}^{M} y_{iJ^{k_{i}}}(m), \forall (i,J^{k_{i}}) \in \mathcal{A} \\
z_{iJ^{k_{i}}} \leq \gamma_{iJ^{k_{i}}}P_{iJ^{k_{i}}}^{k_{i}}, \forall (i,J^{k_{i}}) \in \mathcal{A}  \\
\displaystyle\sum_{k_{i}=1_{i}}^{K_{i}} P_{iJ^{k_{i}}}^{k_{i}} \leq P_{i}, \forall i \in \mathcal{N}.
\end{gather}
where $P_{i}$ is given $\forall$ $i$, $x_{iJ^{k}}^{t_{m}}(m) \in F_{iJ^{k}}^{t_m}(m)$, and $F_{iJ^{k}}^{t_m} (m)$ a bounded polyhedron made of flow conservation constraints:
\vspace{-2mm}
\begin{gather}
\begin{split}
\displaystyle\sum_{(J^{k_{i}}_{l}|(i,J^{k_{i}}_{l})\in \mathcal{A'})} & x_{iJ^{k_{i}}_{l}}^{t_{m}}(m) - \displaystyle\sum_{(J^{k_{i}}_{l}|(J^{k_{i}}_{l},i)\in \mathcal{A'})}x_{J^{k_{i}}_{l}i}^{t_{m}}(m)= s_{i}(m), \\
&\forall i \in \mathcal{N}, \forall t_{m},   \forall m
\end{split} \\
\begin{split}
x_{iJ^{k_{i}}_{l}}^{t_{m}}(m) = \displaystyle\sum_{(J^{k_{i}}_{l} \in J^{k_{i}}|(iJ^{k_{i}})\in \mathcal{A})}x_{iJ^{k_{i}}}^{t_{m}}(m), \\
\forall (i,J^{k_{i}}_{l}) \in \mathcal{A'}, \forall t_{m}, \forall m
\end{split}\\
x_{iJ^{k_{i}}_{l}}^{t_{m}}(m) \geq 0, \forall (i,J^{k_{i}}_{l}) \in \mathcal{A'}, \forall m, \forall t_{m}\in [1,T_{m}].
\vspace{-4mm}
\end{gather}
As opposed to standard multicommodity flow problem in which flows
are simply added over a link, the constraint $(13)$ in fact catches
the essence of network coding by taking only the maximum among all
the flows of a session (note that we only consider intra-session
network coding). Since $F_{iJ^{k_{i}}}^{t_m} (m)$ is the polyhedron
formed by the laws of flow conservation, constraint $(18)$
translates the flow conservation laws from the underlying directed
graph $\mathcal{A'}$ to the hypergraph $\mathcal{A}$ (the wireless
network) by adding the flows on all hyperarcs between node $i$ and
$J_{j}^{k_{i}}$ i.e. flow in $(i,J^{k_{i}}_{j})\in \mathcal{A'}$ is
the sum of all the flows on the hyperarcs $(i,J^{k_{i}})$, $\forall
J^{k_{i}} \ni J^{k_{i}}_{j}$.

As we can see, the above mentioned problem is a convex optimization
problem. The only nonlinear constraint is (13), and could be readily
replaced by the set of linear inequality constraints
$\mathit{y_{iJ^{k_{i}}}(m)} \geq (x_{iJ^{k_{i}}}^{t_{m}}(m))$,
$\forall t_{m} \in [1,T_{m}]$. The modified problem results in a
standard linear multicommodity flow problem with linear objective
and linear constraint set. \vspace{-2mm}
\begin{align*}
minimize \displaystyle\sum_{(i,J^{k_{i}}) \in \mathcal{A}} P_{iJ^{k_{i}}}^{k_{i}}  &&\text{(B)}
\vspace{-4mm}
\end{align*}
subject to:
\vspace{-2mm}
\begin{gather}
\mathit{y_{iJ^{k_{i}}}(m)} \geq (x_{iJ^{k_{i}}}^{t_{m}}(m)), \forall t_{m}, \forall m, \forall (i,J^{k_{i}})\in \mathcal{A} \\
z_{iJ^{k_{i}}}=\displaystyle\sum_{m=1}^{M} y_{iJ^{k_{i}}}(m), \forall (i,J^{k_{i}}) \in \mathcal{A} \\
z_{iJ^{k_{i}}} \leq \gamma_{iJ^{k_{i}}}P_{iJ^{k_{i}}}^{k_{i}}, \forall (i,J^{k_{i}}) \in \mathcal{A}  \\
\displaystyle\sum_{k_{i}=1_{i}}^{K_{i}} P_{iJ^{k_{i}}}^{k_{i}} \leq P_{i}, \forall i \in \mathcal{N}.
\vspace{-4mm}
\end{gather}
where $x_{iJ^{k_{i}}}^{t_{m}}(m) \in F_{iJ^{k_{i}}}^{t_{m}}(m)$, and $F_{iJ^{k_{i}}}^{t_{m}}(m)$ is a bounded polyhedron made of flow conservation constraints.
Note that we optimize the power over each hyperarc, to determine the optimal rates for each hyperarc that satisfies the network demands, we simply need to multiply the optimal power with $\gamma_{iJ^{k_{i}}}$. We will prefer to solve the problem by proposing a decentralized
algorithm for generally understood and appreciated reasons.

\section{DECENTRALIZED ALGORITHM}
\vspace{-2mm}
For developing a decentralized solution for problem $(B)$ we need to
understand the structure of the primal problem first and transform
it into a separable form. We know that the objective function is a
linear and increasing in its domain and so are the constraints.

Taking the Lagrangian dual of the problem $(B)$ we get the dual optimization problem as:
\vspace{-2mm}
\begin{align*}
maximize \big( \displaystyle\sum_{(i,J^{k_{i}}) \in \mathcal{A}} q_{iJ^{k_{i}}} + \displaystyle\sum_{i \in \mathcal{N}} \zeta_{i}P_{i} \big) &&\text{(C)}
\vspace{-2mm}
\end{align*}
subject to:
\vspace{-2mm}
\begin{equation} (\lambda,\mu) \geq \textbf{0}  \end{equation}
\vspace{-2mm}
where,
\vspace{-1mm}
\begin{equation}
\begin{split}
q_{iJ^{k_{i}}} & = q_{iJ^{k_{i}}}(\lambda,\nu,\mu,\zeta,\textbf{x},\textbf{y},\textbf{z}, \textbf{P}) \\
& = \displaystyle\min_{x_{iJ^{k_{i}}}^{t_{m}}(m) \in F_{iJ^{k_{i}}}^{t_{m}}(m)}\bigg(P_{iJ^{k_{i}}}^{k_{i}}+\\
& \displaystyle\sum_{m=1}^{M}\displaystyle\sum_{t_{m}=1}^{T_{m}}(\lambda_{iJ^{k_{i}}}^{t_{m}}(m)) (x_{iJ^{k_{i}}}^{t_{m}}(m) - y_{iJ^{k_{i}}}(m)) +\\
&\nu_{iJ^{k_{i}}}(\displaystyle\sum_{m=1}^{M}y_{iJ^{k_{i}}}(m) - z_{iJ^{k_{i}}})+\\
&\mu_{iJ^{k_{i}}}(z_{iJ^{k_{i}}} - \gamma_{iJ^{k_{i}}}P_{iJ^{k_{i}}}^{k_{i}}) + \zeta_{i}P_{iJ^{k_{i}}}^{k_{i}} \bigg).
\vspace{-2mm}
\end{split}
\vspace{-6mm}
\end{equation}
The dual problem is clearly hyperarc separable and could be solved
in a decentralized manner. But the dual problem is not
differentiable at all the points in the dual domain, this is due to
the fact that there might not be a unique minimizer of
$q_{iJ^{k_{i}}}$ for every dual point as the objective function is a
minimum over sum of linear functions for fixed dual variables. To
solve the dual problem $(C)$, we need to solve its subproblem
$(25)$. The subproblem $(25)$ (and the dual problem $(C)$) could be
solved with a lot of techniques, [10, Chapters 8-10], [11-Chapters
5-6, 12-Chapters 6] using some subgradient based technique but they
do not necessarily yield the primal solution (which is of our
interest here). There are however, methods for recovering primal
solutions from the dual optimizers.

We will take a different technique than the above mentioned
approaches but before lets look into some inter-dependence
characteristics of the dual and primal problem structures. Simply having convex primal problem in hand does not
guarantee strong duality, but with some constraint qualifications we
can assert that strong duality holds or not. One such simple
constraint qualification technique is called \textit{Slater's
condition}.



In our case it can be easily seen for constraints $(13)$ (or $20$)
of problem $(A)$ (or $(B)$), there exist a vector
\textbf{$\{x_{iJ^{k_{i}}}^{t_{m}}(m)\}$} for which the inequality
can be strict.

Let us represent the set of primal vectors as $\textbf{p} =
\{\textbf{x,y,z,P} \} \in S_{1}$ where $S_{1}$ is the feasible set
for the primal problem, and similarly we can do it for the dual
problem, $\textbf{d}=\{\lambda,\nu,\mu,\zeta\} \in S_{2}$. As we can
see that the primal and dual optimal are equal (thanks to strong
duality), we can express our problem in the standard saddle point
form
$\displaystyle\max_{\textbf{d}\in S_{2}} \displaystyle\min_{\textbf{p}\in S_{1}} \phi(\textbf{p,d}) = \displaystyle\min_{\textbf{p}\in S_{1}} \displaystyle\max_{\textbf{d}\in S_{2}} \phi(\textbf{p,d}),$
where function $\phi$ is the Lagrangian dual of the problem $(B)$. This implies that for $(C),$ we get the hyperarc separable saddle-point form
\vspace{-2mm}
\begin{equation}
\displaystyle\max_{\textbf{d}\in S_{2}}q_{iJ^{k_{i}}} = \displaystyle\min_{\textbf{p}\in S_{1}}\displaystyle\max_{\textbf{d}\in S_{2}} \phi(\textbf{p,d}).
\vspace{-2mm}
\end{equation}
Now we are in the position where we can solve the problem, separable
in hyperarcs using any saddle-point optimization method for
non-smooth functions. For our problem set up, we propose a
Primal-Dual Subgradient Algorithm by Nesterov for nonsmooth
optimization [ref. 13]. Nesterov's method generates a subgradient
scheme intelligently based on Dual-Averaging method which beats the
lower case complexity bound for any black-box subgradient scheme.
The algorithm works in both primal and dual spaces, generating a
sequence of feasible points, and ultimately squeezing the duality
gap to zero by finally approaching the optimal solution. A positive
consequence of the Primal-Dual approach is that at each iteration we
get a pair of points $\textbf{(p,d)}$ which are primal and dual
feasible, hence, we get the primal feasible solution with
essentially no extra effort. As opposed to many subgradient type
methods where there needs to be a method for primal recovery,
specially for large and ill-posed problems. \vspace{2mm}
\textit{A. Primal-Dual Subgradient Algorithm.}\\
Since the dual function is hyperarc separable, we can optimize the power
over each hyperarc separately and add each of the optimal solutions
to construct the optimal solution of the dual problem $(C)$, ultimately achieving the primal optimal solution for problem $(B)$. The algorithm is as follows:
\begin{enumerate}
\item Initialization: Set $s_{0}=0 \in Q$. Choose $\theta > 0$.
 \item Iteration $(k \geq 0)$:
    \begin{itemize}
        \item Compute $g_{k} = \partial \phi(p_{k},d_{k})$.
        \item Choose $\sigma_{k} > 0$ and set $s_{k+1}=s_{k}+g_{k}$.
        \item Choose $\theta_{k+1} \geq \theta_{k}$\\
        Set $y_{k+1}=\frac{\theta_{k}}{S_{k}}
        arg\displaystyle\max_{x \in Q} \bigg(\displaystyle\sum_{i=0}^{k} \sigma_{i}\langle g(y_{i}), y_{i}-y \rangle\bigg)$
    \end{itemize}
\end{enumerate}
where $(g_{p},g_{d})$ is the set of primal and dual subgradients and $\sigma_{k}, s_{k}$ and $S_{k}$ are aggregated sequence of points.

\section{SIMULATIONS}
\vspace{-2mm} We now show the results of our simulations that
support the claims of the algorithm presented. We solved the dual
problem in a decentralized way by solving it for every hyperarc
separately and then adding up the respective solutions to construct
the dual optimal solution of the problem $(C)$, when this is
optimal, it is the primal optimal solution for problem $(A)$ in our
case.

The setup consists of uniformly placed nodes on a chosen area of
$a\times a$ $m^{2}$, with given node locations. We start our
simulations with smaller networks of only 4 nodes on a $10 \times
10$ $m^{2}$ area with the area size increasing as the number of
nodes in the network increase to keep the node density/area in a
controlled range. Each node has a single hyperarc and it can
communicate with all the nodes in the network, this is just a simple
generalization of our case where a node can communicate with only a
subset of total nodes in the network. For each network we randomly
choose a set of $m$ multicast sessions and $T_{m}$ set of receivers
for each session respectively with the required rate demand
associated with each session that need to be established, but making
sure the the traffic demands are $\leq$ the respective min-cut for
each session to make the problem feasible.

In Figure 3, we compare the optimal solution approximations of the
Primal-Dual Subgradient Method for problem $(C)$ with the standard
infeasible path following method for problem $(B)$. It can be seen
that the our proposed algorithm gives close approximations of the
primal solution of the problem $(B)$. Note that the path following
method is directly applied to the primal problem and the Primal-Dual
subgradient method is applied to the dual problem, to compute the
dual solution of the problem $(C)$, which will be give us the close
approximation to the primal solution of problem $(A)$.
\begin{figure}
  \begin{center}
    \includegraphics[width=0.46\textwidth]{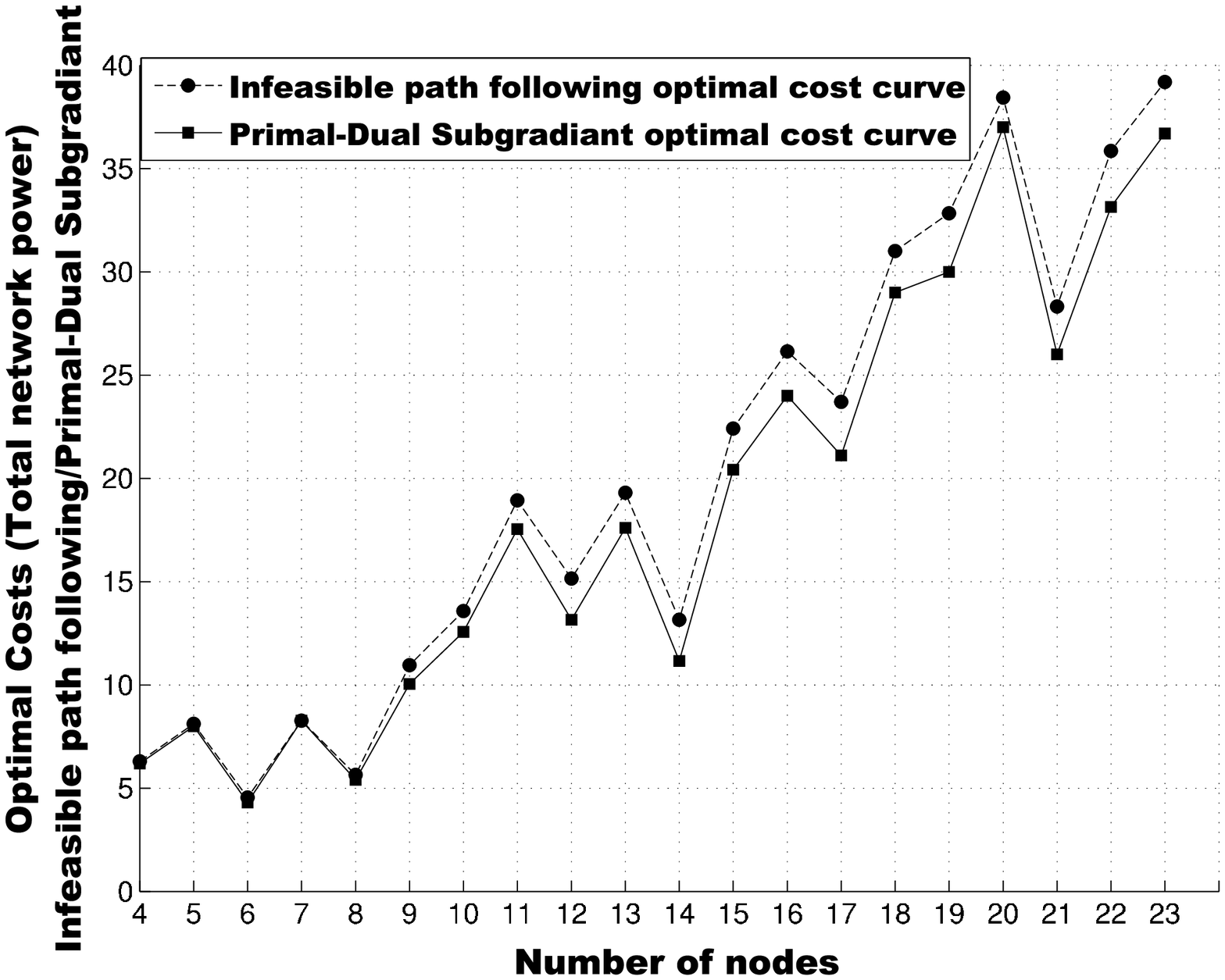}
\vspace{-6mm}
\end{center}
  \caption{Y-axis denotes 2 items, optimal primal costs computed using the infeasible
path following method when applied directly to primal problem $(B)$
and the optimal cost for the dual (primal optimal solution to $(B)$)
when Primal-Dual Subgradient Algorithm is applied to the dual
problem $(C)$.} \vspace{-4mm} \label{Low-SNR Optimal Rate Results.}
\vspace{-2mm}
\end{figure}

\section{CONCLUSION}
\vspace{-2mm} We develop an efficient optimization model that
provides an achievable rate region for the low SNR broadcast channel
and MAC. We do this by showing that rate optimization for the Low
SNR physically degraded broadcast wireless network can be formulated
as a standard linear multicommodity flow problem for optimizing
power over each hyperarc using network coding. Our model is relieved
from interference related issues, this is due to the fact that the
capacity of the low SNR wideband channel is essentially linear in
SNR per for vanishing SNR in the limit, which relieves the system
from interference and related issues. Our model operates in the
non-trivial feasible rate region that achieves capacity in the limit
of disappearing SNR with appropriate encoding scheme.

We use a primal-dual algorithm to construct a decentralized solution
for solving the problem, which has apparent advantages for
recovering the primal solution than standard projected subgradient
methods. In the simulation results shown, we don't present the gains
of routing using network coding over simple routing. But is already
a vast literature establishing this fact.

Finally, we believe that realizing low SNR networks is a worthwhile
attempt as the linearity of capacity in the limit $SNR \rightarrow
0$ provides a fundamental simplicity for networking to be done.
Insights reveal interesting and promising work could be build up and
blended with our simple model (e.g. mobility, reliability etc),
which remains to be explored in this scenario.



\begin{thebibliography}{1}
\vspace{-2mm}
\bibitem{1}
R. S. Kennedy, "Fading Dispersive Communication Channels", New York: Wiley-Interscience, 1969.

\bibitem{2}
E. Telatar, D. N. C. Tse, "Capacity and mutual information of wideband multipath fading channels", IEEE transactions on information theory, vol.46, pp. 1384-1400, July 2000.

\bibitem{3}
Nadia Fawaz, Muriel M\'{e}dard, In Proc in ISIT 2010, "On the
Non-Coherent Wideband Multipath Fading Relay Channel".

\bibitem{4}
R. Ahlswede, N. Cai, S.-Y. R. Li, and R. W. Yeung, "Network information flow", IEEE Trans. Inform. Theory, vol. 46, no. 4, pp. 1204-1216, July 2000.

\bibitem{5}
S.-Y. R. Li, R. W. Yeung, and N. Cai, "Linear network coding", IEEE Trans. Inform. Theory, vol. 49, no. 2, pp. 371-381, Feb. 2003.

\bibitem{6}
R. Koetter and M. M\'{e}dard, "An algebraic approach to network coding", IEEE/ACM Trans. Networking, vol. 11, no. 5, pp. 782-795, Oct. 2003.

\bibitem{7}
D. S. Lun, N. Ratnakar, M. M\'{e}dard, R. Koetter, D. R. Karger, T. Ho, E. Ahmed, and F. Zhao, "Minimum-cost multicast over coded packet networks", IEEE Trans. Inform. Theory, 52(6):2608-2623, June 2006.



\bibitem{8}
L. Zheng, D.N.C. Tse, M. M\'{e}dard, ``Channel Coherence in the Low SNR Regime'', IEEE International Symposium on Information Theory, June 2004. pg. 416.

\bibitem{9}
S. Verd\'{u}, "Spectral efficiency in the wideband regime", IEEE Trans. Info. Theory, 48(6):1319-1343, June 2002.

\bibitem{10}
D. P. Bertsekas, \textit{Network Optimization: Continuous and Discrete Models}, Belmont, MA: Athena Scientific, 1998.

\bibitem{11}
D. P. Bertsekas and R. Gallager, \textit{Data Networks}, 2nd ed. Upper Saddle River, NJ: Prentice Hall, 1992.

\bibitem{12}
D. P. Bertsekas, \textit{Nonlinear Programming}, Belmont, MA: Athena Scientific, 1995.



\bibitem{13}
Yu Nesterov, "Primal Dual Subgradient Methods for convex problems", Mathematical Programming, Volume 120, Issue 1 (April 2009).

\bibitem{14}
R. McEliece and E.C. Posner and L. Swanson,"A Note on the Wideband
Gaussian Broadcast Channel",The Telecommunications and Data
Acquisition Progress 42-84,1985.

\bibitem{15}
Thomas Cover, \textit{Elements of Information Theory}, 2nd ed. Wiley
Series in Telecommunications.


\end{thebibliography}
\end{document}